\documentclass{elsart}
\usepackage{graphicx}
\usepackage{amssymb,amsmath}
\usepackage{graphics}
\usepackage{epsfig}
\journal{Nuclear Physics A}

\def\fmn#1#2{\mbox{${\textstyle \frac{#1}{#2}}$}}
\newcommand{\dd}{\mbox{\rm d}}
\newcommand{\vm}{\mbox{$\phantom{-}$}}

\begin{document}
\begin{frontmatter}
\title{Deuteron analysing powers in deuteron-proton elastic scattering at 1.2 and 2.27~GeV}
\author[1,2] {D.~Mchedlishvili},
\author[1,3] {Z.~Bagdasarian},
\author[4] {S.~Barsov},
\author[3,5] {S.~Dymov},
\author[3] {R.~Engels},
\author[3] {R.~Gebel},
\author[3,4] {K.~Grigoryev},
\author[3,12] {J.~Haidenbauer},
\author[3] {M.~Hartmann},
\author[3] {A.~Kacharava},
\author[3] {I.~Keshelashvili},
\author[6] {A.~Khoukaz},
\author[5] {V.~Komarov},
\author[7] {P.~Kulessa},
\author[5] {A.~Kulikov},
\author[3] {A.~Lehrach},
\author[1] {N.~Lomidze},
\author[3] {B.~Lorentz},
\author[1] {G.~Macharashvili},
\author[3,5] {S.~Merzliakov},
\author[4] {S.~Mikirtychyants},
\author[1] {M.~Nioradze},
\author[3] {H.~Ohm},
\author[6] {M.~Papenbrock},
\author[3] {D.~Prasuhn},
\author[3] {F.~Rathmann},
\author[3] {V.~Serdyuk},
\author[5] {V.~Shmakova},
\author[3] {H.~Str\"oher},
\author[1] {M.~Tabidze},
\author[5] {D.~Tsirkov},
\author[5,8,9] {Yu.~Uzikov},
\author[3,4] {Yu.~Valdau},
\author[10] {C.~Wilkin}\corauth[cor]{Corresponding author}
\ead{c.wilkin@ucl.ac.uk}

\address[1]{High Energy Physics Institute, Tbilisi State University, GE-0186 Tbilisi, Georgia}
\address[2] {SMART$|$EDM-Lab, Tbilisi State University, GE-0179 Tbilisi, Georgia}
\address[3] {Institut f\"ur Kernphysik, Forschungszentrum J\"ulich, D-52425 J\"ulich, Germany}
\address[4] {St.\ Petersburg Nuclear Physics Institute, NRC Kurchatov Institure, RU-188350 Gatchina, Russia}
\address[5] {Laboratory of Nuclear Problems, JINR, RU-141980 Dubna, Russia}
\address[12]{Institute for Advanced Simulation, Forschungszentrum J\"{u}lich, D-52425 J\"{u}lich, Germany}%
\address[6] {Institut f\"ur Kernphysik, Universit\"at M\"unster, D-48149 M\"unster, Germany}
\address[7] {H.~Niewodnicza\'{n}ski Institute of Nuclear Physics PAN, PL-31342 Krak\'{o}w, Poland}
\address[8] {Dubna State University, RU-141980 Dubna, Russia}
\address[9] {Department of Physics, M.~V.~Lomonosov Moscow State University, RU-119991 Moscow, Russia}
\address[10] {Physics and Astronomy Department, UCL, Gower Street, London, WC1E 6BT, UK}

%
%
\begin{abstract}
The vector ($A_y^d$) and tensor ($A_{xx}$ and $A_{yy}$) analysing powers in
deuteron-proton elastic scattering have been measured in the forward
hemisphere at deuteron kinetic energies of $T_d=1.2$~GeV and 2.27~GeV using
the ANKE spectrometer at the COSY storage ring. The results are compared
with other experimental data and with predictions made within the framework
of Glauber multiple scattering theory.
\end{abstract}
\end{frontmatter}
%
%
\section{Introduction}\label{sec.Intro}
\setcounter{equation}{0}

There has been an extensive programme using the ANKE spectrometer at the COSY
Cooler Synchrotron of the Forschungszentrum J\"{u}lich to measure the
analysing powers in small-angle deuteron charge exchange on hydrogen, $dp\to
ppn$~\cite{MCH2013}. At low excitation energies of the final diproton system,
these measurements have yielded valuable information on the spin-spin
amplitudes in large-angle neutron-proton elastic scattering~\cite{BUG1987}.
However, in order to measure such analysing powers, it is first necessary to
determine the polarisation of the incident deuteron beam.

The polarisation of deuteron beams has been studied in several different ways
at COSY~\cite{CHI2006}. The circulating deuteron beam is polarised
perpendicularly to the horizontal plane of the machine. The beam vector
($P_z$) and tensor ($P_{zz}$) polarisations are labeled conventionally in the
reference frame of the source. In contrast, all the spin observables
discussed later refer to the right-handed coordinate system of the reaction
frame, where the beam defines the $Z$-direction while the stable spin axis of
the beam points along the $Y$-direction, which is perpendicular to the COSY
orbit.

The deuteron beam is injected into the COSY ring at 75.6~MeV and the vector
polarisation $P_z$ of the beam is measured at this stage with the low energy
polarimeter (LEP)~\cite{EGG1999}. After acceleration in COSY, the deuteron
vector and tensor ($P_{zz}$) polarisations could be measured in the EDDA
polarimeter~\cite{ALT2005}. Tight limits were placed on any loss of
polarisation during acceleration by measuring in parallel in ANKE the
analysing powers for elastic deuteron-proton scattering, quasi-free
$\pol{n}p\to d\pi^0$, and the $\pol{d}p\to{}^3\textrm{He}\,\pi^0$
reaction~\cite{CHI2006} but subsequent research has shown that the most
efficient tensor polarimeter is based upon the charge exchange
$\pol{d}p\to\{pp\}_sn$ reaction, where only diproton events $\{pp\}_s$ with
low excitation energy are retained~\cite{MCH2013,CHI2006}. The use of this
reaction as a tensor polarimeter was first advocated in Ref.~\cite{BUG1987}.

The methodology for carrying out these measurements has been described at
length in previous publications~\cite{MCH2013} and so the experimental
description given in Section~\ref{sec.Expt} will be relatively brief.
Emphasis is there laid on the polarised source modes used in the experiment
and how the results of the different modes can be combined effectively.
Numerical values of the analysing powers are to be found here. The
model used to interpret the experimental results, which is presented in
Section~\ref{sec.Model}, is based upon the Glauber eikonal
approach~\cite{GLA1955}, which has already been used at a variety of
energies~\cite{TEM2015}.

The results of our measurements of $A_y^d$, $A_{yy}$, and $A_{xx}$ at small
angles at 1.2~GeV and 2.27~GeV are shown graphically in
Section~\ref{sec.Results}, where they are compared with the more extensive
data sets in the 1.2~GeV region from ANL~\cite{HAJ1987} and
SATURNE~\cite{ARV1988}. The results are in general agreement with the
predictions of the modified Glauber model, though the description seems to
get better as the energy increases. The data at 2.27~GeV are especially
important because they were used to determine the deuteron beam polarisations
that were needed in the study of the analysing power in neutron-proton
elastic scattering in the search for a dibaryon~\cite{ADL2014}. Our
conclusions are drawn in Section~\ref{sec.Conc}.

%
%

\section{The experiment}\label{sec.Expt}
\setcounter{equation}{0}

The experiments were carried out at the COoler SYnchrotron (COSY) of the
Forschungszentrum J\"{u}lich using the ANKE magnetic spectrometer, which is
located at an internal target position within a chicane in the storage ring.
Although ANKE contains several detection possibilities, only those of the
Forward Detector were used to measure elastically scattered deuterons or the
two fast protons from the $dp\to\{pp\}n$ charge-exchange reaction. Further
details on the set-up are to be found in the earlier
publications~\cite{MCH2013}.

For this experiment, the COSY polarised source was tuned to produce seven
different combinations of deuteron vector and tensor polarisations plus an
unpolarised beam (\#8), as summarised in Table~\ref{allpol}. The results
shown in the table were obtained using the LEP polarimeter at injection and
ANKE measurements at 1.2~GeV of the deuteron charge-exchange reaction on
hydrogen for $P_{zz}$ and the quasi-free $np \to d\pi^0$ reaction for $P_z$.
It is important to note that, due to its low anomalous magnetic moment, there
are no depolarising resonances for deuteron kinetic energies below 11~GeV
and, as a consequence, it is not expected that deuterons should depolarise
during acceleration at COSY, where the maximum energy is less than 2.3~GeV.
This belief was confirmed by all the ANKE data taken at a variety of
energies~\cite{MCH2013}.

Since the vertical acceptance of the ANKE spectrometer is very small, away
from the forward cone it is not possible to measure the full azimuthal
distribution of any reaction and this makes it difficult to separate effects
arising from the deuteron vector and tensor polarisations. It is therefore
desirable to have beam polarisations that are purely vector or tensor. This
is possible for $P_z$ and mode \#1 is produced with $P_{zz}=0$, which makes
it ideal for the measurement of the vector analysing power in a reaction. On
the other hand, combinations of modes are employed to deliver effectively
``pure'' tensor polarisations and for this more care has to be taken in the
handling of the data shown in Table~\ref{allpol}. This table shows the ideal
values of the polarisations, as requested of the source, as well as values
measured with the LEP and at ANKE.

\begin{table}[h]
\renewcommand{\arraystretch}{1.5}
\centering
\begin{scriptsize}
 \begin{tabular}{c|r|r|c|c|c|c|c}
  \hline
 \#&  $P_{z}^I$ & $P_{zz}^I$ & $P_{z}$ by LEP        & $P_{z}$ by ANKE      & $P_z$ calibrated& $P_{zz}$ by ANKE &$P_{zz}$ calibrated\\
  \hline
  1 & $-\fmn{2}{3}$ &$0$&$  -0.541 \pm 0.008$  & $  -0.522 \pm 0.053$ &$-0.500\pm0.033$ & $\vm0.004 \pm 0.021$ &\\
  \hline
  2 & $+\fmn{1}{3}$&$-1$&$\vm0.197 \pm 0.008$  & $\vm0.307 \pm 0.056$ &$\vm0.223\pm0.024$ & $  -0.548 \pm 0.022$ &\\
  \cline{1-5}
  7 & $-1$&$-2$&$  -0.491 \pm 0.012$  & $  -0.255 \pm 0.057$ &$-0.451\pm0.033$   & $  -0.228 \pm 0.022$ &$-0.443\pm0.020$\\
  \hline
  3 & $-\fmn{1}{3}$&$+1$&$  -0.331 \pm 0.011$  & $  -0.382 \pm 0.053$ &$-0.294\pm0.028$   & $\vm0.498 \pm 0.022$ &\\
  \cline{1-5}
  4 &$0$&$+1$& $\vm0.031 \pm 0.009$  & $  -0.024 \pm 0.056$ &$\vm0.060\pm0.023$ & $\vm0.558 \pm 0.019$ &$\vm0.548\pm0.017$\\
  \hline
  5 &$-1$&$+1$& $  -0.758 \pm 0.007$  & $  -0.746 \pm 0.052$ &$-0.712\pm0.041$   & $\vm0.524 \pm 0.019$ &\\
  \cline{1-5}
  6 &$+1$&$+1$& $\vm0.659 \pm 0.008$  & $\vm0.676 \pm 0.058$ &$\vm0.675\pm0.038$ & $\vm0.411 \pm 0.020$ &$\vm0.465\pm0.017$\\
 \hline
  8 & $0$&$0$&$  -0.007 \pm 0.010$  &  ---                 &  $\vm0.023 \pm 0.023$    & ---&\\
 \hline
 \end{tabular}
 \end{scriptsize}
 \vspace{5mm}
\caption{
Vector and tensor polarisations for eight different configurations of the
polarised deuteron ion source. The ideal values, $P_{z}^I$ and $P_{zz}^I$, are
compared to those obtained using the LEP and ANKE facilities. The methods used
to derive the calibrated values of $P_z$ and $P_{zz}$ are described in the
text.}
 \label{allpol}
\end{table}

The ``calibrated'' values of $P_z$ shown in Table~\ref{allpol} were obtained
by taking the linear fit
\begin{equation}
\label{fit}
P_z(\textrm{ANKE}) =0.030 + 0.979P_z(\textrm{LEP})
\end{equation}
and reading off the values of $P_z$ obtained at the $P_z$(LEP) points. This
corrects some imprecision in the ANKE vector analysing power measurements for
different reactions. The ``calibrated'' tensor values were then obtained by
taking the results of pairs of source modes where the signs of $P_z$ are
opposite but those of $P_{zz}$ are the same. It was then possible to take
linear combinations of the results in order to cancel (within error bars) the
effects of the vector polarisation and the average tensor polarisation for
the pair is given as $P_{zz}$(calibrated). The differences between the values
of $P_{zz}$(calibrated) obtained with the above-mentioned procedure using the
directly measured $P_z$(ANKE) or the ``calibrated'' values of $P_z$ are well
within the quoted error.

The differential cross section for deuteron-proton elastic scattering induced
by a polarised deuteron depends upon the azimuthal angle $\varphi$ of the
recoil deuteron as well as its polar angle $\vartheta$ through the vector
($A_y^d$) and tensor ($A_{xx}$, $A_{yy}$) analysing powers of the reaction,
\begin{eqnarray}
\nonumber
 \frac{\dd\sigma(\vartheta,\varphi)}{\dd\Omega} &=& \left(\frac{\dd\sigma(\vartheta)}{\dd\Omega}\right)_{\!0}
\left[1 + \fmn{3}{2}P_zA_{y}^d(\vartheta)\cos\varphi\right.\\
&&\hspace{1cm}\left.+ \fmn{1}{4}P_{zz}\{A_{xx}(\vartheta)(1-\cos2\varphi)
+A_{yy}(\vartheta)(1+\cos2\varphi)\}\right]\!, \label{polcs}
\end{eqnarray}
where $(\dd\sigma(\vartheta)/\dd\Omega)_{0}$ is the unpolarised cross section
and $P_z$ and $P_{zz}$ are vector and tensor polarisations of the beam,
respectively.

\begin{figure}[hbt]
\includegraphics[width=0.45\columnwidth]{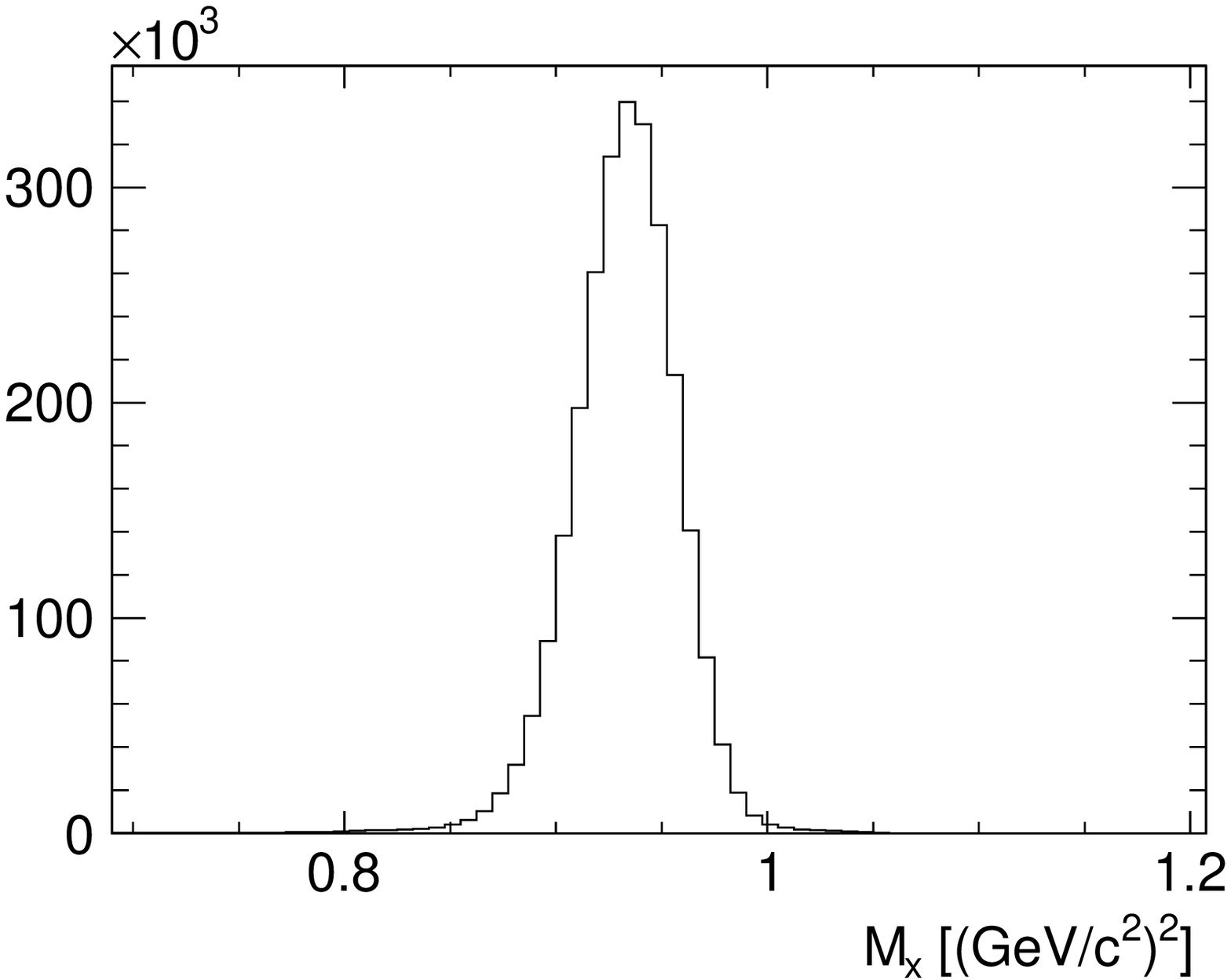}
\hspace{3mm}
\includegraphics[width=0.45\columnwidth]{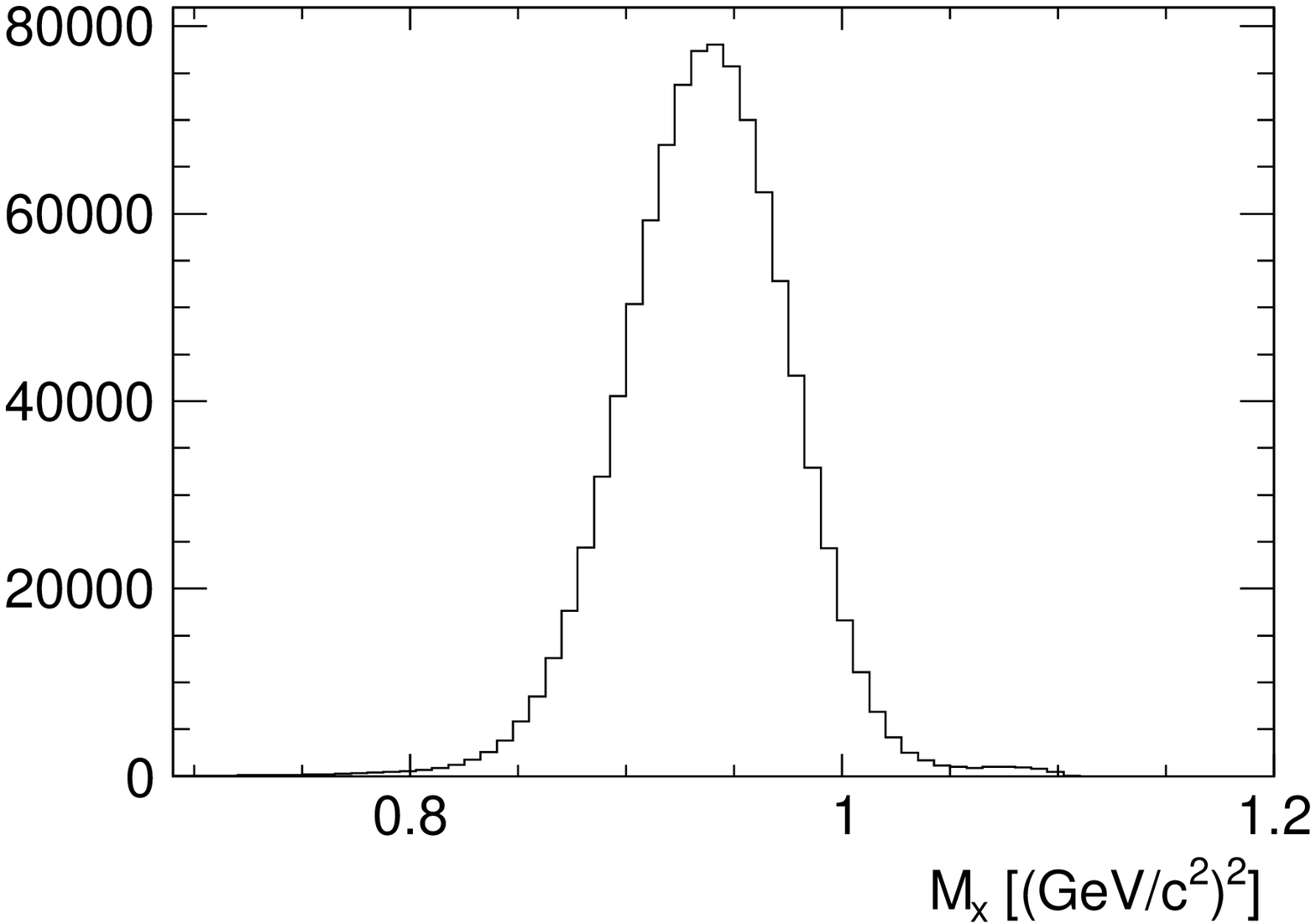}
\caption{\label{fig:Mx} Missing-mass spectra for
the $p(d,d)X$ reaction at 1.2 (left) and 2.27~GeV (right).}
\end{figure}

Deuteron-proton elastic scattering has a very large cross section at small
momentum transfers and the reaction can be identified from the
single-particle momentum spectrum of the fast deuteron measured in the ANKE
Forward Detector (see Fig.~12 in Ref.~\cite{CHI2006}). Triton production
through the $dp\to{}^3\textrm{H}\,\pi^+$ reaction is negligible in
comparison. The corresponding missing-mass spectra (shown in
Fig.~\ref{fig:Mx}) are very clean, with essentially no background, which
confirms the reaction identification. Events lying within a band of $\pm
3\sigma$ of the central proton peak were retained and classified as
corresponding to deuteron-proton elastic scattering.

Values of the vector analysing power $A_y^d$ were obtained by fitting the
normalised $\cos\varphi$ distribution using only polarised mode \#1, where
the tensor polarisation is zero. Numerical results are presented in
Table~\ref{results} in terms of the deuteron c.m.\ scattering angle. They are
also illustrated graphically in Section~\ref{sec.Results} after the
theoretical model has been described. The vector analysing power signal at
2.27~GeV is significantly smaller than at 1.2~GeV, and is also smaller than
that measured at Argonne at 2~GeV~\cite{HAJ1987}. This is consistent with
Argonne data taken at 1.2, 1.6 and 2~GeV, that clearly show a decrease in the
vector analysing power signal as the beam energy is raised.\\

\begin{table}[hbt]
\begin{center}
\begin{tabular}{|c|c|c|c|c|}
\hline
 $T_d$  & $\vartheta_{cm}$& $A_y^d$           & $A_{yy}$        &$A_{xx}$\\
\hline
1.20~GeV&   $16.3^{\circ}$& $0.354\pm0.009$ & $0.204\pm0.017$ & $-0.25\pm0.15$\\
        &   $18.5^{\circ}$& $0.384\pm0.011$ & $0.279\pm0.021$  & $-0.39\pm0.12$\\
        &   $20.4^{\circ}$& $0.400\pm0.011$ & $0.297\pm0.021$  & $-0.36\pm0.15$\\
        &   $22.8^{\circ}$& $0.403\pm0.010$ & $0.365\pm0.022$  & $-0.49\pm0.18$\\
        &   $25.5^{\circ}$& $0.405\pm0.013$ & $0.434\pm0.027$  & $-0.63\pm0.30$\\
        &   $28.5^{\circ}$& $0.399\pm0.015$ & $0.514\pm0.033$  &               \\
        &   $31.7^{\circ}$& $0.350\pm0.019$ & $0.567\pm0.029$  &               \\
\hline
2.27~GeV&   $17.7^{\circ}$& $0.256\pm0.012$ & $0.333\pm0.023$  & $-0.62\pm0.20$\\
        &   $19.7^{\circ}$& $0.263\pm0.016$ & $0.411\pm0.033$  & $-0.68\pm0.19$\\
        &   $21.6^{\circ}$& $0.274\pm0.015$ & $0.482\pm0.032$  & $-0.55\pm0.22$\\
        &   $24.0^{\circ}$& $0.288\pm0.017$ & $0.640\pm0.039$  & $-1.02\pm0.30$\\
        &   $26.6^{\circ}$& $0.281\pm0.025$ & $0.801\pm0.055$  &               \\
        &   $29.4^{\circ}$& $0.210\pm0.036$ & $0.692\pm0.064$  &               \\
        &   $32.5^{\circ}$& $0.045\pm0.045$ & $0.380\pm0.072$  &               \\
\hline
\end{tabular}
\end{center}
\vspace{5mm}
\caption{Vector and tensor analysing powers of the $\pol{d}p\to dp$ reaction
measured at 1.20 and 2.27~GeV in bins centred at the centre-of-mass angles
shown. Note that, due to the much smaller vertical
acceptance than horizontal, the error bars on $A_{xx}$ are significantly bigger
than those on $A_{yy}$. At the larger angles the errors on $A_{xx}$ increase so
much that it is not useful to quote values.
\label{results}}
\end{table}

The tensor analysing powers $A_{xx}$ and $A_{yy}$ were determined from fits
to the normalised $\cos2\varphi$ distributions separately for the three
different pairs of polarisation modes listed in Table~\ref{allpol}. The
results, averaged over these modes, are also given in terms of the deuteron
c.m.\ scattering angle in Table~\ref{results}. Due to the limitations imposed
by the ANKE vertical acceptance, the values of $\varphi$ are always close to
zero so that the precision on the measurements of $A_{xx}$ is much poorer
than for $A_{yy}$. Just as for $A_y^d$, the new results for $A_{yy}$ agree
well with existing data in the 1.2~GeV region. It is seen from the Argonne
data at 1.2, 1.6 and 2~GeV that the zero-crossing point tends to move towards
smaller angles as the energy increases and this is not inconsistent with new
ANKE results at 2.27~GeV. However, some of this effect is kinematic in the
sense that the data are far more stable if presented in terms of the momentum
transfer $t$. Further discussion of the results is deferred until the
theoretical model has been presented in the next section.

%
%

\section{Theoretical model}\label{sec.Model}
\setcounter{equation}{0}

The Glauber single and double-scattering model~\cite{GLA1955} has been used
to estimate the values of the spin observables in proton-deuteron elastic
scattering, following the formalism developed in Ref.~\cite{PLA2010}. This
approach, which includes the full spin dependence of the elementary
proton-nucleon scattering amplitudes and the $S$- and $D$-state components of
the deuteron wave function, allows the calculation of the unpolarised
differential cross section as well as  vector and tensor analysing powers.
This formalism was further developed in the Madison reference frame and
extended to allow the calculation of spin-correlation
parameters~\cite{TEM2015}. Numerical results obtained at the rather low
proton beam energies of 135 - 250~MeV were found to be in reasonable
agreement with existing experimental data in the forward
hemisphere~\cite{TEM2015}.

The $dp \to dp$ transition matrix element can be written as
\begin{eqnarray}
\langle\, p'\mu',d'\lambda'|T|p\mu,d\lambda\,\rangle =
\bar{u}_{\mu'}{\varepsilon_\beta^{\lambda'}}^{\dagger}
T_{\beta \alpha} (\vec{p}, \vec{p'},\vec{\sigma})\,
\varepsilon_\alpha^{\lambda}u_{\mu},
\label{tfi}
\end{eqnarray}
where $u_{\mu}$ ($u_{\mu'}$)  is the Pauli spinor of the initial $p$ (final
$p'$) proton in the state with the spin projection $\mu$ ($\mu'$),
$\varepsilon_\alpha^{\lambda}$ ($\varepsilon_\beta^{\lambda'}$) is the
polarisation vector of the initial $d$ (final $d'$) deuteron in the state
with spin projection $\lambda$ ($\lambda'$), and $T_{\beta \alpha}$ is a
tensor of second rank ($\beta, \alpha$=$x,y,z$), constructed from the momenta
$\vec{p}$ and $\vec{p'}$ and the Pauli matrices $\vec{\sigma}$. After
imposing parity conservation and invariance under rotations, it can be shown
that the transition matrix element of Eq.~(\ref{tfi}) contains 18 invariant
amplitudes, which reduces to 12 once time-reversal invariance is invoked. The
model can therefore also be used to investigate the violation of
time-reversal invariance in nucleon-nucleon interactions~\cite{TEM2016}.

In the  Madison coordinate system, where the $z$-axis lies along $\vec{p}$,
the $y$-axis along $\vec{p}\times \vec{p'}$ and the $x$-direction is chosen
to form a conventional right-handed coordinate system, one can express the
nine elements of $T_{\beta \alpha}$ as
\begin{eqnarray}
\begin{array}{lll}
\hspace{-8mm}T_{xx}=M_1+M_2\sigma_y, & T_{xy}=M_7\sigma_z+M_8\sigma_x,  &T_{xz}=M_9+M_{10}\sigma_y,  \\
\hspace{-8mm}T_{yx}=M_{13}\sigma_z+M_{14}\sigma_x\phantom{a},& T_{yy}=M_3+M_4\sigma_y,  & T_{yz}=M_{11}\sigma_x+M_{12}\sigma_z,   \\
\hspace{-8mm}T_{zx}=M_{15}+M_{16}\sigma_y, & T_{zy}=M_{17}\sigma_x+M_{18}\sigma_z,\phantom{a} & T_{zz}=M_5+M_6\sigma_y,
\end{array}
\label{Txx-Tzz}
\end{eqnarray}
where $\sigma_x$, $\sigma_y$, and $\sigma_z$ are the Pauli matrices, and
$M_i$ ($i=1,\dots, 18$) are the 18 complex amplitudes.

In the work of Platonova and Kukulin~\cite{PLA2010} the $pd\to pd$ transition
amplitude was written in an alternative reference frame, where the $x$ axis
lay along $\vec{q}=\vec{p}-\vec{p'}$ and the $z$-axis along $\vec{k}=
\vec{p}+\vec{p'}$, with the $y$-axis ($\hat{n}$) forming the right-handed
coordinate frame. This choice is dictated by the desire to keep the symmetry
between the initial and final states that has often proved useful in Glauber
theory~\cite{GLA1955}. It also has the added benefit of allowing
time-reversal invariance to be imposed in a much simpler way than is possible
directly in the Madison frame. In order to recast the formulae of
Ref.~\cite{PLA2010} in the Madison frame, one  has  to perform a rotation of
their reference frame around the OY axis through an angle $\theta/2$, where
$\theta$ is the scattering angle, and to make the replacements $OY\to -OY$
and $OX\to -OX$. The 18 amplitudes $M_i$ can be expressed as linear
combinations of the 12 time-reversal-invariant amplitudes $A_i$, which were
derived in  Ref.~\cite{PLA2010}. The corresponding formulae for $M_i$  are
given in Eq.~(5) of Ref.~\cite{TEM2015}.

The vector and tensor analysing powers are expressed in terms of the $M_i$
amplitudes through:
\[A_y^d=
-2Im\Bigl [M_1M_9^*+M_2M_{10}^*+M_{13}M_{12}^*+M_{14}M_{11}^*+M_{15}M_5^*+M_{16}M_6^*\Bigr ]/{\sum_{i=1}^{18}|M_i|^2},\]
\[A_{yy}=1-3\Bigl [|M_3|^2+|M_4|^2+|M_7|^2+|M_8|^2+|M_{17}|^2+|M_{18}|^2\Bigr ]/{\sum_{i=1}^{18}|M_i|^2},\]
\begin{equation}
A_{xx}=1-3\Bigl [|M_1|^2+|M_2|^2+|M_{13}|^2+|M_{14}|^2+|M_{15}|^2+|M_{16}|^2\Bigr ]/{\sum_{i=1}^{18}|M_i|^2}.
\end{equation}


The $pd\to pd$ invariant amplitudes $M_i$ are calculated using the
spin-dependent single and double scattering Glauber model, as described in
Ref.~\cite{TEM2015}. The elementary proton-nucleon elastic scattering
amplitudes are taken in the following form~\cite{PLA2010}:
\begin{eqnarray}
\label{pnamp}
M_N(\vec{p},\vec{q};\vec{\sigma}, \vec{\sigma}_N)=
A_N+C_N\vec{\sigma}\cdot\hat{n} +C_N^\prime\vec{\sigma}_N\cdot\hat{n} +
B_N(\vec{\sigma}\cdot\hat{k})(\vec{\sigma}_N\cdot\hat{k})+\nonumber\\
+ (G_N+H_N)(\vec{\sigma}\cdot\hat{q})(\vec{\sigma}_N\cdot\hat{q})
+(G_N-H_N)(\vec{\sigma}\cdot\hat{n})(\vec{\sigma}_N\cdot\hat{n}),
\end{eqnarray}
where the complex amplitudes $A_N,C_N,C_N^\prime,B_N,G_N,H_N$ ($N=p,n$) have
been calculated from the SAID program~\cite{SAID} and parametrised as sums
of Gaussians.

%

\section{Results}\label{sec.Results}
\setcounter{equation}{0}

The numerical values of our measurements of the deuteron vector and tensor
analysing powers of $\pol{d}p\to dp$ elastic scattering are to be found in
Table~\ref{results}. Here we wish to compare the results with other data and
with the Glauber model predictions in the form of figures.\\[1ex]

\begin{figure}[hbt]
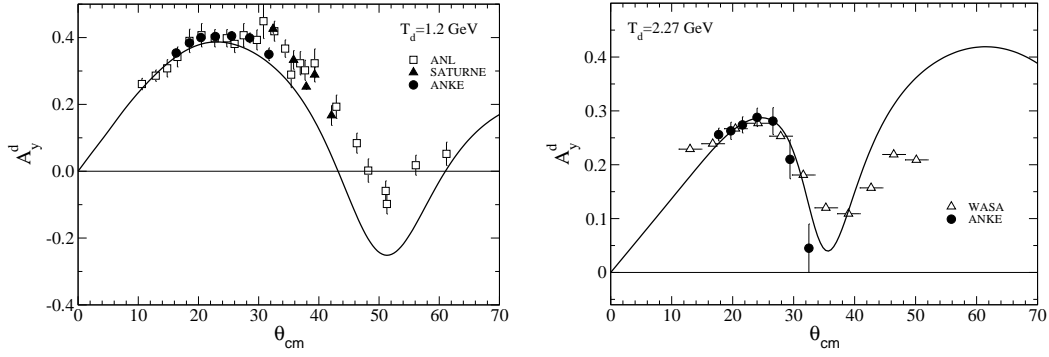

\begin{center}
\includegraphics[width=0.48\columnwidth]{zig2a.eps}
\hspace{2mm}
\includegraphics[width=0.48\columnwidth]{zig2b.eps}
\caption{\label{fig:Ayd} Deuteron vector analysing power $A_y^d$ of elastic
deuteron-proton scattering at 1.2~GeV (left) and 2.27~GeV (right). The ANKE
data (closed circles) at both energies are compared to the
predictions of the extended Glauber model discussed in
Section~\ref{sec.Model}. The SATURNE data at 1.2~GeV~\cite{ARV1988} are shown
as triangles and the ANL data at 1.194~GeV~\cite{HAJ1987} as open squares. The
WASA data at 2.27~GeV~\cite{ADL2014} are shown by open triangles. It should be
noted here that the ANKE results at small angles were used to determine the
beam polarisation in the WASA analysis.}
\end{center}
\end{figure}

As seen in Fig.~\ref{fig:Ayd}, there is good agreement on the deuteron vector
analysing power between the current data and the published experimental
results in the vicinity of 1.2~GeV~\cite{{HAJ1987},{ARV1988}} and at
2.27~GeV~\cite{ADL2014}. Also shown in the figure are curves corresponding
to the predictions of the extended Glauber model of
Section~\ref{sec.Model}.

\begin{figure}[hbt]
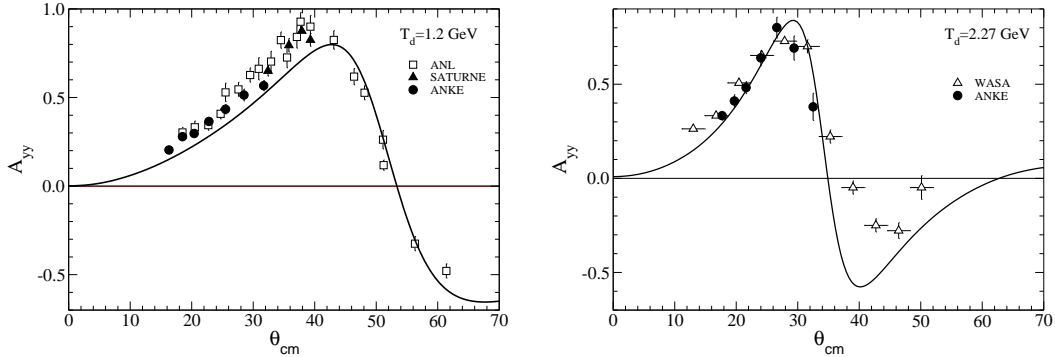

\vspace{0.5cm}
\begin{center}
\includegraphics[width=0.48\columnwidth]{zig3a.eps}
\hspace{3mm}
\includegraphics[width=0.48\columnwidth]{zig3b.eps}
\caption{\label{fig:Ayy} The results and predictions for the tensor analysing
power $A_{yy}$ using the same notation as that employed for the vector
analysing power in Fig.~\ref{fig:Ayd}.}
\end{center}
\end{figure}

The analogous results and calculations for the tensor analysing power
$A_{yy}$ are to be seen in Fig.~\ref{fig:Ayy}. Of particular note here is the
behaviour of the WASA data which, though they show a dip at
$\theta_{cm}\approx 40^{\circ}-45^{\circ}$, this is not as deep as in the
model which, as shown in Ref.~\cite{PLA2010}, reproduces the dip seen in the
ANL data at the neighbouring energy of 2.0~GeV~\cite{HAJ1987}.

\begin{figure}[hbt]
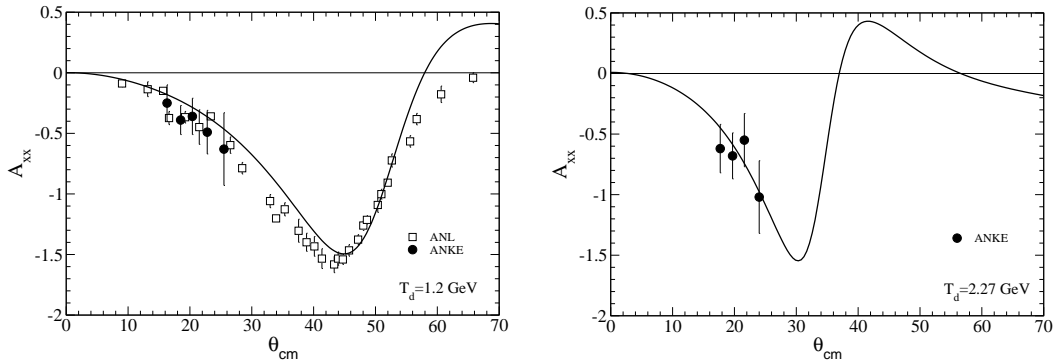

\vspace{5mm}
\begin{center}
\includegraphics[width=0.48\columnwidth]{zig4a.eps}
\hspace{3mm}
\includegraphics[width=0.48\columnwidth]{zig4b.eps}
\caption{\label{fig:Axx} ANKE measurements (black circles) of the tensor
analysing power $A_{xx}$ at 1.2~GeV (left) and 2.27~GeV (right) are compared
with the ANL data at 1.2~GeV~\cite{HAJ1987} (open squares) and to the
predictions made within the framework of the extended Glauber model of
Section~\ref{sec.Model}.}
\end{center}
\end{figure}

Finally in Fig.~\ref{fig:Axx} we compare the current measurements of $A_{xx}$
at 1.2~GeV and 2.27~GeV with the model predictions. As previously stressed,
the limited acceptance of the ANKE exit window in the $y$ direction means
that the errors on $A_{xx}$ are significantly larger than those shown for
$A_{yy}$. Nevertheless, the data do follow the trends of the theoretical
curves.

%
%
\section{Conclusions}\label{sec.Conc}
\setcounter{equation}{0}

We have presented new measurements of the deuteron analysing powers $A_y^d$,
$A_{yy}$, and $A_{xx}$ in small-angle deuteron-proton elastic scattering at
$T_d=1.2$~GeV and 2.27~GeV. The values of $A_y^d$ and $A_{yy}$ obtained at
the lower energy are completely consistent with those derived earlier at
ANL~\cite{HAJ1987} and SATURNE~\cite{ARV1988}. It should be noted that at the
higher energy there is agreement by construction with the WASA
results~\cite{ADL2014} at the smallest angles because the WASA data were
normalised to those of ANKE in this region. However, the $A_y^d$ data may
start diverging at the largest angles.

We have also shown the results of an extended single- plus double-scattering
Glauber model~\cite{TEM2016} that has proved very useful at small angles.
By comparing its predictions with the ANL data at 1.2, 1.6, and
2.0~GeV~\cite{HAJ1987}, it seems that the approach becomes more reliable as the
energy increases. The model reproduces the main features of the ANKE
data and this brings into question the WASA data at
2.27~GeV, which seem to have a quantitatively different behaviour to the
model in the dip regions of both $A_y^d$ and $A_{yy}$.

Though the error bars on the ANKE values of $A_{xx}$ are understandably quite
large, the data are not in conflict with the model though the 1.2~GeV data on
both this and $A_{yy}$ would even fall on the theoretical curves with a
slight increase in beam tensor polarisation. However, this possibility seems
unlikely in view of the ANL and SATURNE data. On the other hand, ANKE
measurements of various  spin observables in deuteron charge exchange on
hydrogen, and subsequent theoretical analysis, suggest that there might be
improvements to the SAID neutron-proton charge exchange amplitudes at
1.135~GeV~\cite{MCH2013}. This energy is, of course, close to the limit of
the SAID neutron-proton predictions~\cite{SAID}.

%
%
\section*{Acknowledgements}
We are grateful to the accelerator crew for the reliable operation of COSY
and the deuteron polarimeters. The work was supported in part by the COSY FFE
programme, the Heisenberg-Landau programme, and the Shota Rustaveli National
Science Foundation Grant 09-1024-4-200.


\end{document}